\documentclass[a4paper]{jpconf}
\usepackage{lineno}
\usepackage{graphicx}
\usepackage{hyperref}
\usepackage{url}
\usepackage[utf8]{inputenc}

\begin{document}

\title{Optimizing CMS build infrastructure via Apache Mesos}

\author{David Abdurachmanov$^1$, Alessandro Degano$^2$, Peter Elmer$^3$, Giulio
Eulisse$^1$, David Mendez$^4$, Shahzad Muzaffar$^1$ on behalf of the CMS collaboration.}

\address{$^1$ Fermilab, Batavia, IL 60510, USA}
\address{$^2$ Università di Torino}
\address{$^3$ Department of Physics, Princeton University, Princeton, NJ 08540,
USA}
\address{$^4$ Universidad de los Andes, Bogotá, Colombia}

\ead{Giulio.Eulisse@cern.ch}

\begin{abstract}

The Offline Software of the CMS Experiment at the Large Hadron Collider (LHC) at
CERN consists of 6M lines of in-house code, developed over a decade by nearly
1000 physicists, as well as a comparable amount of general use open-source code.
A critical ingredient to the success of the construction and early operation of
the WLCG was the convergence, around the year 2000, on the use of a homogeneous
environment of commodity x86-64 processors and Linux.

Apache Mesos is a cluster manager that provides efficient resource isolation and
sharing across distributed applications, or frameworks. It can run Hadoop,
Jenkins, Spark, Aurora, and other applications on a dynamically shared pool of
nodes.

We present how we migrated our continuous integration system to schedule jobs on
a relatively small Apache Mesos enabled cluster and how this resulted in better
resource usage, higher peak performance and lower latency thanks to the dynamic
scheduling capabilities of Mesos.

\end{abstract}

\section{Current build infrastructure}

CMS Core Software group is responsible for the whole release engineering process
for the CMS Offline Software, CMSSW. In particular this includes:

\begin{itemize}

\item Release building and packaging, for both production releases and bi-daily
integration builds for production and experimental branches alike.

\item Unit testing and reduced release validation for integration builds.

\item Automated Testing and Continuous Integration of updates as provided by
various collaborators under the form of Github Pull Requests.

\item Running additional web based QA and documentation tools.

\item Deployment of nightly integration builds via CVMFS to the collaboration.

\end{itemize}

CMS builds up to 21 releases per day on Scientific Linux 6 (SLC6) alone,
including production releases, development ones and experimental branches. In
the last month ( April 7, 2015 – May 7, 2015) we had over 294 active pull
requests tested (eventually multiple times) and integrated in our release
branches and integration a total of over 8,000 pull requests have been processed
by our Continuous Integration machinery since we started using Github in the
middle of 2013.~\cite{Eulisse2014}

To manage handling the automated QA and release workflows, the CMS build
infrastructure currently consists of a cluster of 284 virtual cores and 540 GB
of memory provisioned in the CERN OpenStack based cloud.~\cite{CERNOPENSTACK}

The QA procedures, the Continuous Integration is performed in an automatic manner
via Jenkins~\cite{JENKINS}, an OpenSource continuous integration system.

Moreover the group is responsible for maintaining a number of web services,
mostly serving build artifacts like installation RPMs, log files and ROOT files
resulting from the tests.

\section{Old, statically partitioned, build infrastructure architecture}

In the past in order to handle the load we could simply statically partition our
OpenStack quota to have a good balance between the size of the build cluster and
the size of the services. Moreover we provisioned build machines which were big
enough for building a release with high level of parallelism, in order to reduce
the build latency, but still small enough to avoid wasting to much CPU time when
Jenkins scheduled non parallel jobs on them, wasting their potential.

Moreover due to the fact that it's not possible to cross compile SLC5 releases
on SLC6, in the past some of the resources need to be assigned to SLC5 builds,
which results in a net loss of efficiency since SLC5 builds are occasional but
needed to be delivered with the same latency of SLC6 ones so we do not have much
margin in overcommitting the SLC5 infrastructure.

While the static partitioning of the cluster has been sufficient for a long
period, we wish to improve both the latency for  critical builds (i.e.
development builds or tests on pull requests for  development builds) and
improve the overall cluster utilization, with the aim  of reducing the number of
resources needed.

\section{Apache Mesos: a dynamic approach to cluster partitioning}

The latency and resource utilization problems we are facing are actually a
common issue in the industry, where static partitioning of resources is the
typical cause of resource overcommitting and the following low cluster
utilization.

Big names in the industry (e.g. Google~\cite{BORG}) have long solved the problem
with proprietary solutions which provide more dynamic cluster scheduling,
however recently a number of open-source projects have emerged trying to provide
the same kind of functionality. In particular we decided to evaluate a resource
manager called Apache Mesos~\cite{MESOS}(just Mesos from now on).

Mesos originated from research work done at UC Berkeley~\cite{MESOSPAPER} which
has then been brought to production by Twitter. As the project evolved it got
incubated as part of the Apache Foundation Incubator, successfully graduating
from it in 2013~\cite{GRADUATION}, and it's therefore now distributed as
flagship project under the Apache License. It is commonly used by other big
internet companies like Apple, eBay, Vimeo, and universities like UCSF and UC
Berkeley.

Mesos cluster architecture consists of two sets of nodes, masters which
coordinate the work, and slaves which execute it. In particular Mesos is a so
called two-level scheduler: the slave nodes offer their available resources to
masters which are then responsible for handling them to the so called frameworks
standalone software components which are responsible for the application
specific scheduling policy. When proposed a resource offer, a framework can
decide whether to accept it or not, and in case it does it can specify which and
how many tasks it would like to create for the given resource offer. This allows
for more fine tuned scheduling of resources and for application aware scheduling
and has been demonstrated in very large scale (thousands of nodes) setups.

On the backend, Mesos masters uses Apache ZooKeeper (just ZooKeeper from now on)
~\cite{ZOOKEEPER} to provide a fault tolerant, distributed, setup when multiple
masters are used.

Moreover, recent improvements allow Mesos to take advantage of a relatively
modern Linux feature, known as \textit{containers}, to isolate the payload task
being executed by the slaves. Containers are middle ground entities between
processes and virtual machines. Like processes, they run all on top of the same
linux kernel, however they are completely isolated one from the other, having
each their own separate resources and runtime. In particular one of the most
popular engines supported by Mesos is Docker~\cite{DOCKER}, which is quickly
gaining industry acceptance and whose usage CMS itself is
exploring~\cite{DOCKERPAPER}.

A number of Mesos frameworks are available and open-sourced, in particular a
simple PaaS called Marathon~\cite{MARATHON}, a plugin which allows two spawn
build nodes on demand for the Jenkins continuous integration system
~\cite{MESOSJENKINS} and a number of frameworks to dynamically provision
resources for Big Data stacks like Apache Hadoop~\cite{MESOSHADOOP}, Apache
Spark~\cite{MESOSSPARK} or databases like Apache Cassandra~\cite{MESOSCASSANDRA}
or Elasticsearch~\cite{MESOSELASTIC}.

\section{New cluster architecture}

We decided to redesign our build infrastructure to take advantage of Mesos and
Docker.

The new cluster uses only two static partitions which from which from now on we
will refer to as masters and slaves. The machine belonging to those two
partitions are provisioned as virtual machines in CERN OpenStack
infrastructure~\cite{CERNOPENSTACK} and their configuration and deployment is
handled using the CERN Puppet / Foreman setup~\cite{CERNCFG}.

\subsection{Masters}

The masters run on three separate virtual machines, each positioned in a
different OpenStack availability zone, in a high-availability setup, so that we
can gracefully handle the loss of one machine, with no service interruption,
which is considered enough for our purpose.

Each master runs the following services:

\begin{itemize}

\item \textbf{Mesos master service}: Mesos masters are responsible for
scheduling services on the rest of the cluster.

\item \textbf{ZooKeeper}: ZooKeeper is used by Mesos to keep track of the
distributed state of the masters, effectively providing high availability to the
system.

\item \textbf{The Marathon PaaS}: Marathon is responsible for scheduling
services running on our cluster, dynamically choosing their placement based on
the provided set of rules and constraint. It acts as a cluster level
\textit{init} or \textit{upstart} daemon.  By using either Marathon GUI or its
REST interface an administrator can request to start a process or a Docker
container, eventually in multiple copies, and Marathon will take care of
starting and monitoring such a process(es), ensuring it's restarted in case it
dies. Moreover it exposes the location of the various services via a REST based
API so that the front-end proxies can use it for service discovery and redirect
traffic to it. A custom made wrapper script is used to generate the front-end
configuration.

\item \textbf{HAProxy}: HAProxy is a fast and easy solution for load balancing
proxy which we use as a front-end to redirect to our services. A cronjob running
every few minutes is responsible for reading the backend information and update
the configuration so that various requests are redirected to the correct host in
the slave cluster.

\item \textbf{nginx}: since HAProxy does not support SSL termination, in case we
need it, we use the nginx web service for this purpose, however, even when this
is required, we leave redirection part to HAProxy.

\end{itemize}

\subsection{Slaves}

The slaves cluster is comprised of nodes running the Mesos slave client and on
which Docker was previously deployed via Puppet. They are responsible for
running the tasks scheduled by the Mesos masters.

In general tasks are running inside a Docker container, which actually allows us
to separate the configuration of the slaves from the configuration of the
application running on it. This means for example that we can run SLC5 Jenkins
builders inside a Docker container which runs on top of a SLC6 machine,
eliminating the special need for a SLC5 build cluster.

In line of principle each node is equal to the other and this is generally true
for the case in which the application does not need to store a large state for a
long period of time. This is in particularly true for Jenkins builders, which
are expendables. For applications which do have a long standing state, e.g. an
Elasticsearch cluster or a web server, we make sure that the machine(s) which
hosts the data has a special attribute set, so that Mesos can easily schedule
the service to the correct backend. An alternative which will soon be possible
is to use Mesos dynamic resources extension which will allow to specify not only
job needs in terms of CPU or memory, but also persistent disk storage.

For security reasons, each node cannot talk directly to each other, but has to
go through the front-end interface.

In its current incarnation our Mesos cluster setup uses the OpenStack based CERN
Cloud for provisioning machines however we have plans to experiment with real
hardware, at least for the stateless builders, since bare metal seem to provide
better performance than standard build machines due to the fact building a large
project like CMSSW is an I/O intensive tasks due to a large number of files
being compiled and linked.

\section{Notable Mesos frameworks and their utilizations}

\subsection{Provisioning slaves with Jenkins Mesos plugin}

Given the fact that CMS Build Infrastructure heavily relies on Jenkins to drive
the release integration and QA process, one of the main advantages of Mesos is
the availability of a Jenkins plugin which allows provisioning new slaves as a
required. In particular the plugin allows to define multiple queues, each with a
separate label associated and each with different requirements in terms of CPU
and memory usage. Together with the fact that Jenkins allows programmatic
definition of labels attached to a job, this allows fine grained scheduling
where, for example, more recent releases are assigned to a queue with a larger
share of the CPU budget. Moreover the ability to assign a Docker container to a
queue allows reducing the issues deriving from build machines being deployed at
different times and therefore potentially having slightly different setups.

\subsection{Deploying services with Marathon and Docker}

Mesos would not be to much different from a normal batch-system, however, if it
did not provide the ability to execute long running services, either running
natively inside a Docker container. In particular the Mesos framework Marathon
provides a simple setup where services can be defined via a JSON formatted file,
passed to the service via its REST API. The JSON file contains information like
which executable to launch or alternatively which Docker container to spawn.
Additional configuration options allow to define port and volume mappings for
the application, and to specify how many copies are requested. Moreover Marathon
allows the service manager to specify placement constraints so that, for
example, a given instance of a web server is started on a machine where the data
is actually located or it allows to horizontally scale an application by forcing
all the instances to run on a different server. For example this is actively
used in our setup to launch a Elasticsearch database running on a three node
cluster where we store build logs and information. The whole cluster
configuration, including the actual location of spawn services on the cluster
and their port mappings, are available via Marathon REST API and can therefore
be used for service discovery and automated configuration. Finally, in order to
improve the experience of deploying a new service Marathon it is possible to
navigate deployment history and roll back to previous configurations. The
flexibility of the system is demonstrated by the fact it's easy to actually
deploy a new Mesos and Marathon cluster using Mesos and Marathon itself. This is
in particular an interesting feature which could be used to provide very dynamic
opportunistic usage of resources by launching and reshaping Mesos Clusters on
top of a parent stable cluster and providing an entry-point to each cluster to
different users.

\section{Advantages of the new setup}

We believe that from the architectural point of view, the new setup has mainly
three advantages:

\begin{itemize}

\item Clear separation between infrastructure and application deployment.

\item Painless redundancy and high-availability.

\item Larger nodes.

\end{itemize}

Separation of concerns in particular improves reliability of the whole
infrastructure during upgrades. By having dockerized applications running on the
cluster we are not in the need of updating them as we update the infrastructure
itself, following CERN/IT upgrades pace. Moreover, high-availability allows us
to update one component at the time, without having to sacrifice service uptime.
Finally the fact we can dynamically use part of a slave allows us to reshape our
cluster to have virtual machines with more cores with the consequent higher
performance in case of parallel workloads, without sacrificing resources in case
of sequential ones.

\section{Conclusions and future work}

We have successfully setup a Mesos cluster and used it to launch jobs via
Jenkins and to deploy services via Marathon. While the system does comes with a
non trivial learning curve and it's still an evolving product, Mesos has proven
 itself to be a reliable method to unify cluster resources and provides high
availability.

Work continues in migrating our build infrastructure to Mesos, in particular
to migrate larger jobs to it.

Given the fact Mesos is specifically thought as a way to have multiple clients
access the same cluster resources, future work can also be expected in the
direction of sharing resources of the build infrastructure cluster with other
parties within and outside CMS, establishing collaboration, for example, with
the build and Release and QA team of other experiments. In order to do so an
economic model of how the cluster resources are shared among competing parties
would be required.

\section*{Acknowledgements}
This work was partially supported by the National Science Foundation, under
Cooperative Agreement PHY-1120138, and by the U.S. Department of Energy.

\section*{References}
\bibliographystyle{unsrt}
\bibliography{summary}

\end{document}